\documentclass[%
 reprint,
 superscriptaddress,
 amsmath,amssymb,
 aps,
pre,
]{revtex4-1}

\usepackage {atbegshi}
\usepackage [dvipdfmx]{graphicx}% Include figure files
\usepackage{dcolumn}% Align table columns on decimal point
\usepackage{bm}% bold math
\usepackage{mediabb}
\usepackage{braket}
\usepackage{multirow}
\usepackage {color}
\begin{document}
\newcommand\red[1]{\textcolor{red}{#1}}
\newcommand\blu[1]{\textcolor{blue}{#1}}

% Use the \preprint command to place your local institutional report
% number in the upper righthand corner of the title page in preprint mode.
% Multiple \preprint commands are allowed.
% Use the 'preprintnumbers' class option to override journal defaults
% to display numbers if necessary
%\preprint{}
\preprint{APS/123-QED}

%Title of paper
\title{Dynamical Crossover in a Stochastic Model of Cell Fate Decision}

% repeat the \author .. \affiliation  etc. as needed
% \email, \thanks, \homepage, \altaffiliation all apply to the current
% author. Explanatory text should go in the []'s, actual e-mail
% address or url should go in the {}'s for \email and \homepage.
% Please use the appropriate macro foreach each type of information

% \affiliation command applies to all authors since the last
% \affiliation command. The \affiliation command should follow the
% other information
% \affiliation can be followed by \email, \homepage, \thanks as well.

\author{Hiroki Yamaguchi}%
\email{yamaguchi@noneq.c.u-tokyo.ac.jp}
\affiliation{%
 Department of Applied Physics, University of Tokyo, 7-3-1 Hongo, Bunkyo-ku, Tokyo 113-8656, Japan
  }%
\author{Kyogo Kawaguchi}%
%\email{Second.Author@institution.edu}
  \affiliation{%
 Department of Systems Biology, Harvard Medical School, Boston, MA 02115, USA
 %This line break forced with \textbackslash\textbackslash
}%
\author{Takahiro Sagawa}%
%\email{Second.Author@institution.edu}
\affiliation{%
 Department of Applied Physics, University of Tokyo, 7-3-1 Hongo, Bunkyo-ku, Tokyo 113-8656, Japan
 %This line break forced with \textbackslash\textbackslash
}%

%\collaboration{MUSO Collaboration}%\noaffiliation

%\author{Charlie Author}
% \homepage{http://www.Second.institution.edu/~Charlie.Author}
%\affiliation{
% Second institution and/or address\\
% This line break forced% with \\
%}%
%\affiliation{
% Third institution, the second for Charlie Author
%}%
%\author{Delta Author}
%\affiliation{%
% Authors' institution and/or address\\
% This line break forced with \textbackslash\textbackslash
%}%

%\collaboration{CLEO Collaboration}%\noaffiliation

%Collaboration name if desired (requires use of superscriptaddress
%option in \documentclass). \noaffiliation is required (may also be
%used with the \author command).
%\collaboration can be followed by \email, \homepage, \thanks as well.
%\collaboration{}
%\noaffiliation

\date{\today}

\begin{abstract}
We study the asymptotic behaviors of stochastic cell fate decision between proliferation and differentiation. We propose a model of a self-replicating Langevin system, where cells choose their fate (i.e. proliferation or differentiation) depending on local cell density. Based on this model, we propose a scenario for multi-cellular organisms to maintain the density of cells (i.e., homeostasis) through finite-ranged cell-cell interactions.
Furthermore, we numerically show that the distribution of the number of descendant cells changes over time, thus unifying the previously proposed two models regarding homeostasis: the critical birth death process and the voter model.
Our results provide a general platform for the study of stochastic cell fate decision in terms of nonequilibrium statistical mechanics.
\begin {description}
% insert suggested PACS numbers in braces on next line
\item[PACS numbers] 05.65.+b, 87.17.Ee, 87.18.Hf
%May be entered using the \verb+\pacs{#1}+ command.
% insert suggested keywords - APS authors don't need to do this
%\keywords{}
\end {description}
\end{abstract}

%\maketitle must follow title, authors, abstract, \pacs, and \keywords
\maketitle

% body of paper here - Use proper section commands
% References should be done using the \cite, \ref, and \label commands
%\section{}
% Put \label in argument of \section for cross-referencing
%\section{\label{}}
%\subsection{}
%\subsubsection{}

%%
\section {Introduction \label {Sec:Introduction}}
\begin {figure} [tbp]
		\begin {center}
		(a)
		\begin {tabular} {c}
		\begin {minipage}{0.5\textwidth}
			%\centering
			\hspace {-0.2in}
			\includegraphics[keepaspectratio, scale = 1.0]{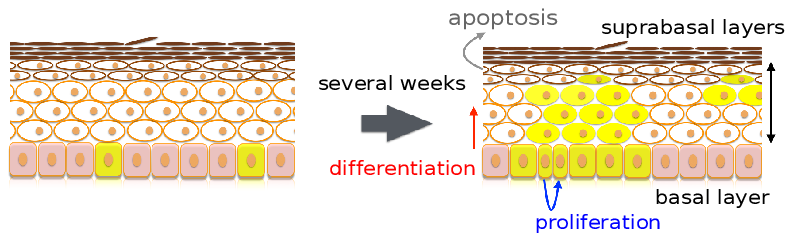}
			\\(b)
		\end {minipage}\\
		\begin {minipage}{0.5\textwidth}
			\centering
			\includegraphics[keepaspectratio, scale=1.0]{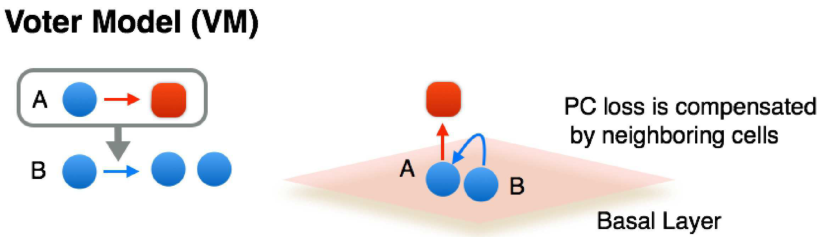}
			\\(c)
		\end {minipage}\\
		\begin {minipage}{0.5\textwidth}
			\centering
			\includegraphics[keepaspectratio, scale=1.0]{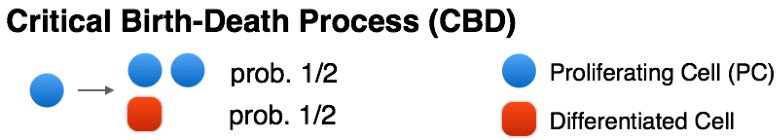}
			\
		\end {minipage}

		\end {tabular}
		\end {center}
		\caption {(Color online) (a)Schematic showing a cellular label tracking experiment. Proliferating cells (pink) confined to the basal layer of the tissue undergo proliferation within the basal layer, as well as differentiation toward the upper layers. Cell fate is tracked by monitoring the labeled clones (yellow), which either expand or shrink by proliferation and differentiation, respectively. (b) Schematic of VM. (c) Schematic of CBD. \label {Fig:Clonal Analysis} }
	\end {figure}

A variety of biological phenomena have been extensively investigated in light of modern nonequilibrium physics. Tissue turnover in multicellular organisms is an interesting example of stationary nonequilibrium system (see Fig.~\ref {Fig:Clonal Analysis}(a)). Throughout adult life, biological tissues are constantly renewed by newly born cells from stem cell pools. The production of cells must be balanced with the death of old cells, which is called tissue homeostasis. Tissue stem cells, which are able to proliferate (i.e., divide) and differentiate into tissue specific cells, play a key role in tissue turnover. Since differentiated cells exit from the proliferation cycle and are eventually killed, tissue homeostasis is maintained by balanced kinetics of stem cell fate (proliferation or differentiation). Intercellular interactions in cell fate decision kinetics are considered to be crucial during homeostasis, and therefore, nonequilibrium statistical mechanics of many-body systems~\cite{Hinrichsen2000, Odor2004} is expected to play an important role to clarify the mechanism of tissue homeostasis.

Recent advances in experiments have enabled the tracing of cell fate dynamics (i.e., kinetics of proliferation and differentiation) in adult mammalian tissues~\cite {Simons2011}. In these experiments, cells in the tissue are labeled by fluorescent proteins, which are inherited by their progeny. Starting from isolated labeled single cells in the basal layer tissue, the cells can divide and expand its population within the basal layer, or get excluded from the basal layer through differentiation (Fig.~\ref {Fig:Clonal Analysis}(a)). The measured population of the labeled cells that survived in the basal layer (i.e., clone) showed scaling behavior in statistics.
Most significantly, the average number of cells in surviving clones showed power-law growth $n_{\rm surv} (t) \sim t^{\delta}$, and the cumulative clone size distribution $C_n(t)$ (i.e., the probability of having a clone with no less than $n$ cells) showed the scaling law $C_n (t) \sim \Phi (n / n_{\rm surv} (t) )$, which both depended on the spatial dimension of the tissue~\cite {Klein2011}. The scaling behavior ruled out one of the classical pictures of tissue homeostasis, where stem cells always undergo asymmetric division (one daughter cell is differentiating and the other maintains its stemness).

The quantitative modeling approach has revealed two canonical examples of stochastic dynamics that explain the scaling behaviors. Importantly, these two models reflect the different mechanisms of cell fate regulation~\cite {Klein2011}. Firstly, cell-extrinsic regulation results in the scaling form with $\delta = 1/ 2$ and $\Phi (X) = e^{-\pi X^2 / 4}$ in one-dimension~\cite {Klein2011, Lopez-Garcia2010, Klein2010}, and $\delta = 1$ and $\Phi (X) = e^{-X}$ in two-dimension~\cite {Klein2011, Sawyer1976, *Bramson1980}. 
These scaling forms are derived from the voter model (VM)~\cite {Holley1975, Dornic2001}, in which a differentiating cell leaving the basal layer is assumed to directly trigger the proliferation of a neighboring cell to compensate for the loss (the mechanism can be vice versa, Fig.~\ref {Fig:Clonal Analysis}(b)).
The second scheme is the cell-intrinsic regulation~\cite {Klein2011, Clayton2007, *Klein2007}, which is described by the critical birth-death process (CBD)~\cite {Galton1873, *Watson1875, *Kendall1966, *Harris2002} (Fig.~\ref {Fig:Clonal Analysis}(c)). This model assumes that a cell stochastically chooses proliferation or differentiation with equal probability independent from other cells, which results in the scaling form with $\delta = 1$ and $\Phi (X) = e^{-X}$~\cite {Clayton2007, *Klein2007}. The extrinsic model (i.e., $\delta = 1/ 2$ and $\Phi (X) = e^{-\pi X^2 / 4}$) was consistent with clonal labeling experiments in some one-dimensional tissues such as intestinal crypts~\cite {Lopez-Garcia2010} and seminiferous tubules~\cite {Klein2010}. On the other hand, the experimental works in skin tissues (i.e., two-dimensional tissues) \cite {Mascre2012, Rompolas2016} showed that both cell-intrinsic and extrinsic models are consistent with the clonal dynamics.

In this paper, we propose a model of cell fate decision, focusing on the cell-cell interaction associated with a finite range. In our model, the population of cells is regarded as a self-replicating many-body Langevin system, where we incorporate intercellular interaction in the self-replication process through local cell density~\cite {Shraiman2005, Ranft2010}, where density is defined with a certain length scale. 
Both short- and long-range interaction can be realized in real tissues as consequences of different regulatory mechanisms. For example, recent experimental~\cite {Montel2011, Coste2010, Dupont2011, Eisenhoffer2012} and theoretical~\cite {Shraiman2005, Ranft2010, Hannezo2014} works suggest that mechanical cues could be relevant in cell fate decision. Long-range interaction via autocrine signaling is also crucial in skin stem cells~\cite {Lim2013}. Therefore, exploring cell fate decision processes from the point of view of cell-cell interaction range would be significant.

We find that homeostasis is maintained in this model as a consequence of the interaction, meaning that global cell density is autonomously kept constant on average. Furthermore, we show that the previously proposed VM and CBD scenarios are incorporated in our model as the small and large limits of the cell-cell interaction range. This indicates that the interaction range of the density-dependent replication process is a key in determining which model of the two appears in biological tissues. We find that in the case of the intermediate value of the interaction range, the clone size statistics cross over from the CBD statistics to the VM statistics as time evolves. We propose that by evaluating the timing of the crossover in experiment, we can indirectly infer the interaction range of the fate decision dynamics in the tissue. Our results also reveal a natural scenario in which VM can arise in real experimental setups.

This paper is organized as follows. In Section \ref {Sec:Model}, we describe our model, based on cell density dependent interaction with a finite interaction range. In Section \ref {Sec: Main results} we give the main results, showing the dynamical crossover of clone size statistics in the case of finite interaction range. In Section \ref {Sec:Scaling hypothesis}, we discuss the nature of crossover, and show the scaling hypothesis for the crossover.

\section {Model \label {Sec:Model}}
\begin {figure} [tbp]
		\begin {center}
		(a)
		\begin {tabular} {c}
			\begin{minipage}{0.5\textwidth}
			\centering 
			%\subcaption { \label {Fig:Configuration} }
			\includegraphics[keepaspectratio, scale=1.0]{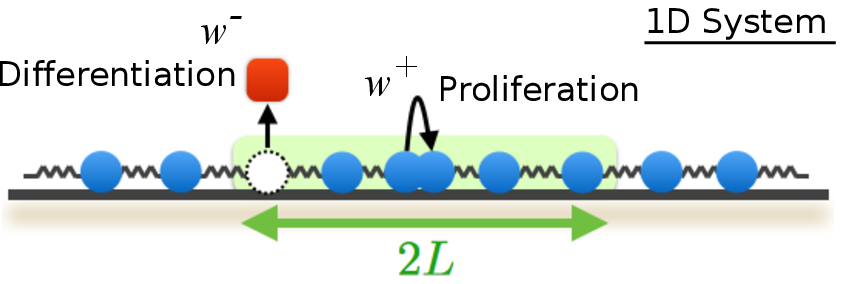}
			\\ \hspace {-1.5in} (b) \hspace {1.5in} (c)
		\end{minipage}
		\\
		\begin{minipage}{0.25\textwidth}
			\centering
			%\subcaption { \label {Fig:DensityDependence} }
			\includegraphics[keepaspectratio, scale=1.0]{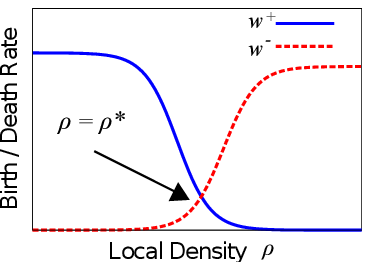}
		\end{minipage}
		\hspace {-0.1in}
		\begin{minipage}{0.25\textwidth}
			\centering
			%\subcaption { \label {Fig:PhaseDiagram} }
			\includegraphics[keepaspectratio, scale=1.0]{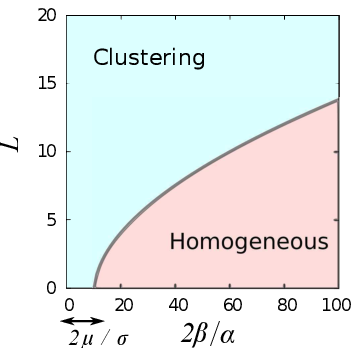}
 		\end{minipage} 
		
		 \end {tabular}
		 \end {center}
		 \caption {(Color online) (a) Schematic of the self-replicating Langevin system. (b) An example of the density dependence of the birth/death rates. (c) The phase diagram via structure factor analysis, which is detailed in Appendix A. The gray line separates the parameter space into the linearly stable and unstable regions. Here, $\alpha$ represents the sensitivity of the growth rate around the steady-state density, and $\beta$ is the effective diffusion constant that depends on the strength of the cell-cell adhesion $K$. $\mu / \sigma$ is the relative noise strength. The definitions of $\alpha$, $\beta$, $\sigma$ and $\mu$ are given in Eq.~\eqref {SupEq:Def_alpha_beta}. Our clonal analysis is performed in the spatially homogeneous region. \label {Fig:Models} }
	\end{figure}

We model the population of stem cells as an interacting many-particle system with $ \{ x_k (t) \}_{k=1} ^{N (t)} $ being the position of the center of the $N (t)$ cells existing on the basal layer at time $t$ (see Fig.~\ref {Fig:Models}(a)). 
We assume for simplicity that the basal layer is occupied only by a single type of stem cell which can move and divide. The irreversible differentiation of the cells is described by the stochastic exclusion of a cell from the dynamics. Considering two or more types of cells coexisting in the basal layer as in the case of previous models~\cite {Clayton2007, Klein2007} will not change the scaling behavior discussed in the following. 
Although our model can be extended to higher dimensions, we here focus on the one-dimensional case in order to study the distinct limits of asymptotic behavior. Neighboring cells in a tissue are typically attached to each other by cell-cell adhesion. We incorporate this interaction by the following many-body Langevin equations~\cite {Puliafito2012}:
\begin {equation}
	\frac {d}{dt} x_j (t) = - \frac {\partial}{\partial x_j} U ( \{ x_k \} ) + \xi_j (t).
	\label {Eq:EOM}
\end {equation}
Here, $ \xi_j (t) $'s represent the white Gaussian noise satisfying
\begin {equation}
	\left < \xi_j (t) \right > = 0, ~ \left < \xi_j (t) \xi_k (s) \right > = 2 D \delta_{j, k} \delta (t - s)
\end {equation}
for $ j, k= 1, \dots, N (t)$. Initial positions of the cells are prepared so that the label of the cells are ordered as $x_{k-1} \leq x_k \leq x_{k+1}$, and the periodic boundary condition is employed. A positive constant $D$ is the noise strength and $ U ( \{ x_k \} ) = \sum_{ k } u ( \lvert x_k - x_{k+1} \rvert ) $ denotes the two-body interaction, which describes the cell-cell adhesion between nearest neighbors. For simplicity, we set $u ( X ) =K \left ( X - l_0 \right )^2 / 2$.
The strength of the adhesive potential $K$ determines a typical time scale for spatial relaxation, and the natural length $ l_0 $ determines a typical length scale of a cell. 

A crucial point of our model is that the tissue homeostasis is achieved as a consequence of the cell-cell interaction. To this end, we assume that the local cell density affects the cell fate decision process~\cite {Shraiman2005, Ranft2010}. We define the local cell density $\rho_L (x)$ as:
\begin {equation}
	\rho_L (x) := \frac {1}{ 2 L } \int_{ x - L}^{x + L} d y \sum_{j} \delta ( y - x_j ), 
	\label {Eq:LocalCellDensity}
\end {equation}
where $L$ denotes the interaction range of cell-cell interactions. The interaction range $L$ corresponds to the biologically relevant length scale of the cell fate regulation through local cell density. For example, the case of long range interaction $L \gg l_0$ can be realized by external chemical control or autocrine signaling. On the other hand, $L \simeq l_0$ describes the situation where the fate is associated with the distance of a cell from its closest neighbors, corresponding for instance to mechanical regulation. A proliferating cell (PC) at position $x$ undergoes the following birth-death process with rates $w^{\pm}$ that depend on local cell density (see Fig.~\ref {Fig:Models}(a)):
\begin {equation}
	PC \stackrel{ }{\longrightarrow}
	\begin {cases}
		PC + PC & \hspace {0.2in} {\rm with~rate~~} \lambda w^+ ( \rho_L (x) ) \\
		\emptyset & \hspace {0.2in}{\rm with~rate~~} \lambda w^- ( \rho_L (x) ),
	\end {cases}
	\label {Eq:Division Kinetics}
\end {equation}
where $\emptyset$ denotes differentiation (i.e., removal from the basal layer), and the typical timescale is set by $\lambda^{-1}$.
 We here do not explicitly assume any correlation between the fates of the two daughter cells. This means that the newly born PCs will independently choose division or differentiatiation according to Eq. \eqref {Eq:Division Kinetics}, as opposed to the asymmetric division scenario where the fates of siblings are strictly anti-correlated.~\cite {Klein2011}. 

We assume that $ F (\rho) := w^+ (\rho) - w^- (\rho) $ has an attractive fixed point $\rho = \rho^*$ such that $ F (\rho^*) = 0$ and $ F^{\prime} (\rho^*) < 0 $, where $ F^{\prime} (\rho) := d F (\rho) / d \rho$. An important role of the attractive fixed point $\rho = \rho^*$ is to regulate cell fate through local cell density in an autonomous fashion (see Fig.~\ref {Fig:Models}(b)). We will see in our model that the power law and scaling law in the clone size statistics appear at the fixed point of cell density, corresponding to the situation where homeostasis is achieved.

Equations \eqref {Eq:EOM} are discretized and numerically solved by the Euler-Maruyama method. The cell fate decision process \eqref {Eq:Division Kinetics} is implemented as follows. When a cell undergoes proliferation, the local cell density $\rho_L$ is evaluated for newly born cells, and two lifetimes $\tau_\pm$ are generated from exponential distributions with rates $\lambda w^\pm (\rho_L)$, respectively. If $\tau^+ < \tau^-$, the cell undergoes proliferation after time $\tau^+$, and otherwise it undergoes differentiation after time $\tau^-$. In order to study the asymptotic clone size statistics, we introduce the label degree of freedom to cells. In our numerical simulations, we initially prepare only one labelled cell and many other unlabeled cells, mimicking the induction of marker protein in experiments. The quantities of interest are the average clone size of the labeled progenies $l (t) := l_0 n_{\rm surv} (t)$ and the labeled clone size distribution $C_n (t)$. The clones here refer to the labeled descendants within the stem cell population. 

We remark on the stability of the model. When $L$ is sufficiently large, the population of cells tends to form spatial clustering. This is regarded as an example of the Brownian bug problem, which has been observed in various models with self-replication and diffusion~\cite {Young2001, *Hernandez-Garcia2004, *Heinsalu2010, *Ramos2008}. We identified the parameter regimes as shown in Fig.~\ref {Fig:Models}(c), in which the clustering of cells does not occur, by analyzing the structure factor (see Appendix A). This argument ensures that our numerical analysis is performed in this homogeneous region. 	
\section {Main results \label {Sec: Main results}}
\begin{figure*} [htbp]
		\begin {center}
		\hspace {-1.95in} (a) \hspace {3.1in} (b) 
		\begin {tabular} {c}
		\begin {minipage}{1.0\textwidth}
		\begin {minipage}{0.475\textwidth}
			\centering
			%\subcaption { \label {Fig:Crossover} }
			\includegraphics[keepaspectratio, scale=1.0]{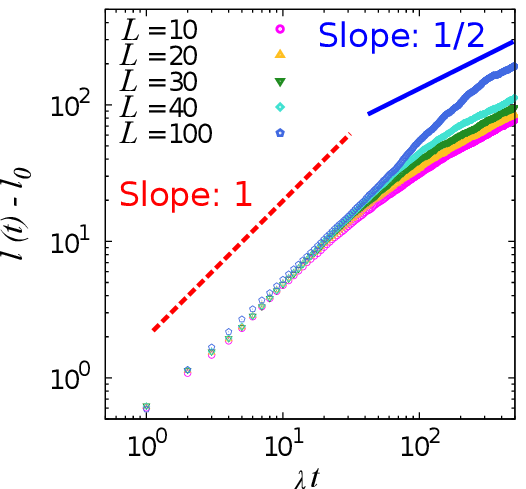}
			\\ \hspace {-2in} (c)
		\end {minipage}
		\begin{minipage}{0.475\textwidth}
			\centering
			%\subcaption { \label {Fig:Crossover_CSD} }
			\includegraphics[keepaspectratio, scale=1.0]{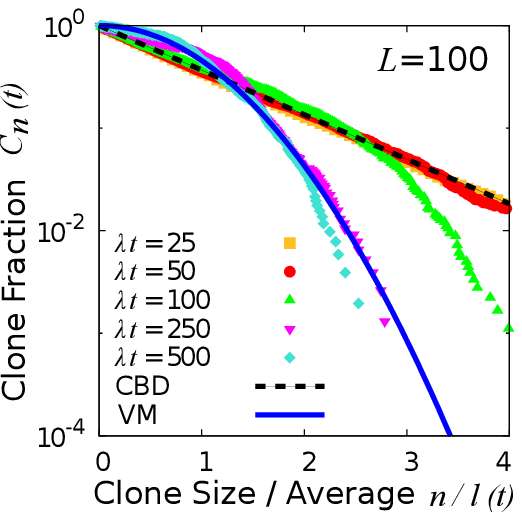}
			\\ \hspace {-2in} (d)
 		\end{minipage} 
		\end {minipage}\\
		\begin {minipage}{1.0\textwidth}
		\begin {minipage}{0.475\textwidth}
   			\centering
			%\subcaption { \label {Fig:CSD_Infinite_inset} }
			\includegraphics[keepaspectratio, scale=1.0]{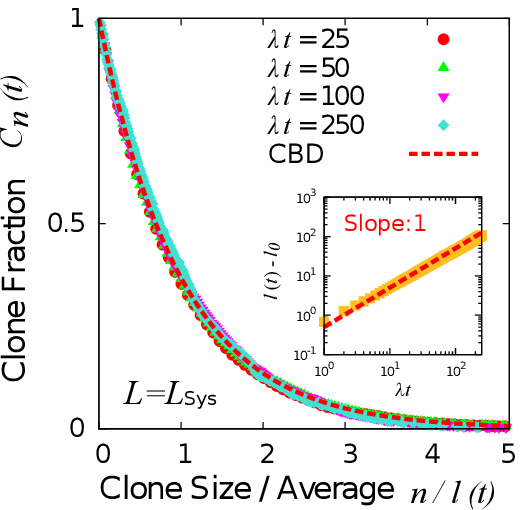}
		\end {minipage}
		\begin {minipage}{0.475\textwidth}
   			\centering
			%\subcaption { \label {Fig:CSD_1_inset} }
			\includegraphics[keepaspectratio, scale=1.0]{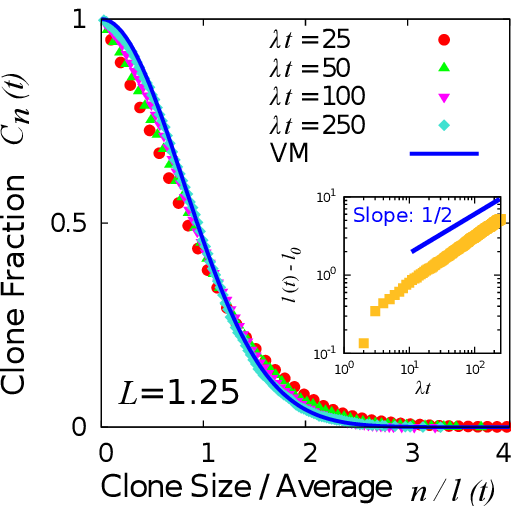}
		\end {minipage}
		\end {minipage}
		 \end {tabular}
		 \caption {(Color online) Numerical results of the asymptotic clone size statistics in clonal analysis. We take $l_0 = 1$ and $L_{\rm Sys} = 1000$. (a) The average clone size $l(t)$ for $L / l_0 = 10, 20, 30, 40$, and $100$. (b) The clone size distributions $C_n (t)$ at different time points for $L / l_0 = 100$ plotted against $ n / l (t)$. (c) (d) $C_n (t)$ plotted against $n / l (t)$ for the large $L$ case ($L = L_{\rm Sys}$) and the small $L$ case ($L / l_0 = 1.25$), respectively. The insets show $l (t) - l_0$ against time $\lambda t$. \label {Fig:Results} } 
		 \end {center}
	\end{figure*}

We now discuss our numerical results. Figure~\ref {Fig:Results}(a) shows the time evolution of the average clone size for several values of $L$. The average clone size grows linearly in the short time scale, and exhibits the power-law growth with exponent $1 / 2$ in the long time scale. Figure~\ref {Fig:Results}(b) shows that the clone size distribution is the exponential form in the short time, and then crosses over to the half-Gaussian form in the long time scale. These results imply that the clone size statistics cross over from the CBD statistics with $\delta = 1$ and $\Phi (X) = e^{-X}$ to the VM statistics with $\delta = 1/2$ and $\Phi (X) = e^{-\pi X^2 / 4}$, in the course of time. 
	
In order to clarify the dynamical crossover of the clone size statistics, we consider the two opposite limits of the interaction range $L$. Since the interaction range $L$ lies between the cell size $l_0$ and the system size (i.e., the size of the tissue) $L_{\rm Sys}$, we call the two limits $L \to l_0$ and $L \to L_{\rm Sys}$ the small $L$ limit and the large $L$ limit, respectively. In our simulations, we take $l_0 = 1$ and $L_{\rm Sys} = 1000$. Figures~\ref {Fig:Results}(c) and~\ref {Fig:Results}(d) show the time evolution of the clone size distributions for $L = L_{\rm Sys}$ and $L = 1.25 l_0$, respectively. The inset in Fig.~\ref {Fig:Results}(c) or~\ref {Fig:Results}(d) shows the time evolution of the average clone size. The clone size statistics asymptotically approach the CBD statistics in the large $L$ limit and to the VM statistics in the small $L$ limit. These results imply that the CBD statistics and the VM statistics are formulated in a unified view through the interaction range $L$.
	
In the following, we consider the mechanism that gives rise to the CBD statistics and the VM statistics in the large and small $L$ limit, where the existence of the attractive fixed point $\rho = \rho^*$ plays a significant role. In the large $L$ limit, the cell fate regulation is governed by global cell density $\rho (t)$ in the continuum limit as $L_{\rm Sys} \to \infty$:
	\begin {equation}
		\lambda^{-1} \frac {d}{d t} \rho = \left ( w^+ (\rho ) - w^- ( \rho ) \right ) \rho = F \left ( \rho \right ) \rho.
	\end {equation}	
The existence of an attractive fixed point $\rho^*$ of $F (\rho)$ ensures homeostasis so that $w^+ (\rho^*) = w^- (\rho^*)$ holds in the long time scale. Therefore, the clone size statistics are expected to behave asymptotically as the CBD statistics: $\delta = 1$ and $\Phi (X) = e^{-X}$. We emphasize that in our model, the critical clone size statistics are achieved as a result of cell-cell interactions through cell density, in contrast to CBD, where the criticality is assumed a priori. The asymptotic CBD statistics can now be interpreted as a consequence of long-range density feedback interaction.

On the other hand, in the case of small $L$, the cell-cell interaction is effectively short-ranged, because the ever-expanding average size of the surviving clones $l (t)$ is always larger than $L$. In this case, as soon as a proliferating cell undergoes differentiation, the proliferation rate increases locally around that position, and the neighboring cells will be likely to compensate for the loss of the adjacent cell. Therefore, the resulting clone size statistics are expected to asymptotically behave as the VM statistics with $\delta = 1/2$ and $\Phi (X) = e^{-\pi X^2 / 4} $. In other words, our model implies that the VM statistics can naturally arise as a result of short-range interaction.

\section {Scaling hypothesis \label {Sec:Scaling hypothesis}}
\begin {figure} [tbp]
	\begin {center}
%		\begin {tabular} {c}
%		\begin {minipage}{0.475\textwidth}

		\includegraphics[keepaspectratio, scale=1.0]{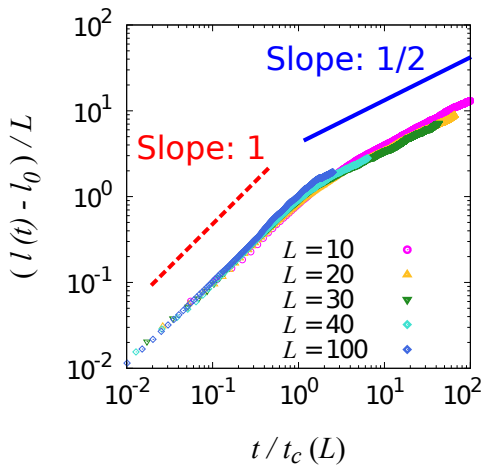}
%		\end {minipage}\\
%		\begin {minipage}{0.475\textwidth}
%		\vspace {0.2in}

%		\includegraphics[keepaspectratio, scale=0.17]{CoarseGraining.eps}
%		\end {minipage}
%		\end {tabular}		
		\caption {(Color online) Scaling form of the average clone size $l (t)$. $l (t) - l_0$ is scaled by $L$ and plotted against $t / t_c (L)$ for $L / l_0 = 10, 20, 30, 40$, and $100$. \label {Fig:CrossoverScaling} }
	\end {center}
\end {figure}

Our numerical results suggest that the time scale for the crossover increases with the interaction range $L$. The dynamical crossover takes place due to the competition between the interaction range $L$ and the average clone size $l (t)$. Since the average clone size is ever-increasing in time, one expects that $l (t)$ exceeds $L$ at certain time $t_c$. We refer to $t_c$ as the crossover time, at which the behavior of the clone size statistics change. Since the clone size statistics are effectively the CBD statistics in short time scale, the crossover time $t_c$ is estimated from the following equation:
\begin {equation}
	L =  l (t_c) \simeq l_{\rm CBD} (t_c) = l_0 \left (1 + \frac {1}{2} \lambda t_c \right ),
\end {equation}
where the average clone size $l (t)$ is approximated by the exact expression of that of the CBD statistics: $l_{\rm CBD} (t) = l_0 (1 + \lambda t / 2)$. Therefore, we expect that $t_c (L) = 2 \lambda^{-1} (L - l_0) / l_0$. Scaling the time by the crossover time $t_c (L)$, and scaling the average clone size $l (t)$ by the interaction range $L$, all curves collapse onto a single master curve, as shown in Fig.~\ref{Fig:CrossoverScaling}. 

From Fig.~\ref {Fig:CrossoverScaling}, we find that the average clone size $l (t)$ has the following scaling form:
\begin {equation}
	\frac { l (t) - l_0 }{L} =  f \left ( \frac {t}{t_c (L)} \right ) 
	\begin {cases}
		\simeq \displaystyle { \frac {t}{t_c (L)} } 						&\hspace {0.05in} t \ll t_c (L), \\
		\sim \displaystyle { \left ( \frac {t}{t_c (L)} \right )^{1/2} }			&\hspace {0.05in} t \gg t_c (L),
	\end {cases}
	\label {Eq:Scaling}
\end {equation}
which reconfirms the CBD statistics and the one-dimensional VM statistics.

\begin {figure} [tbp]
	\begin {center}
		\includegraphics[keepaspectratio, scale=1.0]{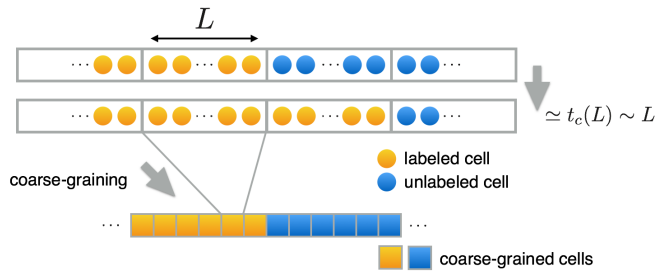}
		\caption {(Color online) Schematic showing the coarse-graining of cell population. After coarse-graining, the clonal dynamics in the long time scale is similar to the case of small $L$ limit. \label {Fig:CoarseGraining} }
	\end {center}
\end {figure}

We now discuss the nature of crossover in the case of $l_0 \ll L \ll L_{\rm Sys}$. The dynamical crossover takes place due to the competition between two length scales: the interaction range $L$ and the ever-expanding average clone size $l (t)$. In the short time scale with $l (t) \ll L$, the cell fate regulation is effectively governed by global cell density, leading to the CBD statistics. On the other hand, the clonal dynamics in the long time scale with $l (t) \gg L$ can be explained by coarse-graining the cell population (see Fig.~\ref {Fig:CoarseGraining}). In the coarse-graining scheme, the total system is divided into a chain of boxes with width $L$, which is the cluster size of cells that essentially feel the same cell density according to Eq.~\eqref{Eq:LocalCellDensity}. 
Now we focus on the two boxes at the boundary between labeled and unlabeled population. Since the clonal dynamics within each box approximately follows the CBD dynamics, the average time, in which all the unlabeled cells in a box are replaced by the labeled cells in the neighboring box (or vice versa) is approximately $t_c (L)$. Therefore, by scaling $t$ by $t_c (L)$ and $x$ by $L$, the clonal dynamics in terms of the coarse-grained local cell density would be similar to that in the small $L$ limit, and thus the VM statistics appear in the long time scale. Note that this result is not implying that the single-cell level kinetics switches from CBD to VM; the VM scaling can only be probed at length scales larger than $L$ and time scales larger than $t_c(L)$, and in shorter length scales or time scales, the dynamics will always appear as CBD.

From the scaling form of $l (t)$ (Eq.~\eqref{Eq:Scaling}), we find that the interaction range $L$ in a tissue can be estimated from $L = l_0 (1 + \lambda t_c / 2)$, where $t_c$ is obtained by fitting experimental data to the $l (t)$ curve. 
Since the crossover appears after the average clone size $l (t)$ reaches the interaction range $L$, the system size needs to be large enough compared with $L$ in order to detect the crossover in finite size tissues.

Our crossover detection scheme would be applicable to one-dimensional systems such as seminiferous tubules. Previous clonal labeling experiments on seminiferous tubules have shown that one-dimensional VM statistics appear in the millimeter order length scale in the time scale of several months~\cite {Klein2010, Klein2011}. 
From the viewpoint of our model, this means that there could be a length scale $L$ that is smaller than a millimeter within which the cell fate dynamics will look more like the autonomous model. Thus, by examining the statistics of smaller clones at shorter time scale, it may be possible to detect the crossover before the clonal dynamics converging to the VM statistics, allowing us to estimate the finite interaction range.

\section {Concluding remarks \label {Sec:Concluding remarks}}
We presented a novel model of stochastic cell fate decision, based on cell-cell interactions through local cell density. In our model, two previous stochastic models (i.e., CBD and VM) are unified by introducing the interaction range $L$. We numerically studied the asymptotic clone size statistics for the one-dimensional case. The asymptotic clone size statistics of CBD and VM are realized in the large and small limits of $L$, respectively. In the case of intermediate $L$, the clone size statistics cross over from that of CBD to that of VM in the course of time. Furthermore, we studied the scaling hypothesis for the dynamical crossover of the average clone size. The one-dimensional tissue experiments revealed the VM statistics without its detailed mechanism, and our study clarified a natural scenario behind the emergence of the VM statistics in biological systems.

Although the mechanism of tissue homeostasis has not yet been revealed at the level of molecular biology, our phenomenological analysis has quantitatively elucidated the role of the time and length scales of cell-cell interactions. The density dependent mechanism of cell fate regulation could arise via either mechanical cues from surrounding cells, external cues from the niche, or long-range autocrine signaling. Extracting the length scale from the crossover would possibly be a key to investigate the nature of regulation in cell fate decision.

We expect that our analysis would provide a platform for further studies. For instance, the scenario of cell fate decision in two-dimensional tissues is still unsettled, since CBD and VM have the same asymptotic statistics, apart from the logarithmic correction in the average clone size~\cite {Klein2011}. On the basis of our model, it is left for future studies to discuss the spatial correlation of labeled cell configuration to clarify the cell fate decision scenario in sheet tissues. 
\begin{acknowledgments}
We are grateful to Allon M. Klein and Kazumasa A. Takeuchi for fruitful comments. TS is supported by JSPS KAKENHI Grant No. 25800217 and No. 22340114, by KAKENHI
No. 25103003 “Fluctuation \& Structure", and by Platform for Dynamic Approaches to Living System from MEXT, Japan.
\end{acknowledgments}

\appendix 
\section {Stability analysis of spatial clustering} %{Linear stability analysis}
In this appendix, we discuss the stability of the spatially uniform distribution of cells in our model. In our numerical simulation, we have observed that the population of cells exhibits spatial clustering, when the interaction range $L$ is sufficiently large. Similar phenomena have been reported in a variety of systems, which has been known as the Brownian bug problem~\cite {Young2001, Hernandez-Garcia2004, Ramos2008, Heinsalu2010}. In the following, we give a simple scenario of the spatial clustering, and clarify the parameter region in which the spatial clustering does not occur.% and the uniform distribution is linearly stable. 
%\end{abstract}

% insert suggested PACS numbers in braces on next line
%\pacs{}
%% insert suggested keywords - APS authors don't need to do this
%%\keywords{}
%
%%\maketitle must follow title, authors, abstract, \pacs, and \keywords
%\maketitle

% body of paper here - Use proper section commands
% References should be done using the \cite, \ref, and \label commands
%\section{}
% Put \label in argument of \section for cross-referencing
%\section{\label{}}
%\subsection{}
%\subsubsection{}

%%
%\section{Setup}
\subsection {Microscopic equation of motion}
\begin {figure} [htbp]
	\begin {center}
		\includegraphics[keepaspectratio, scale=1.0]{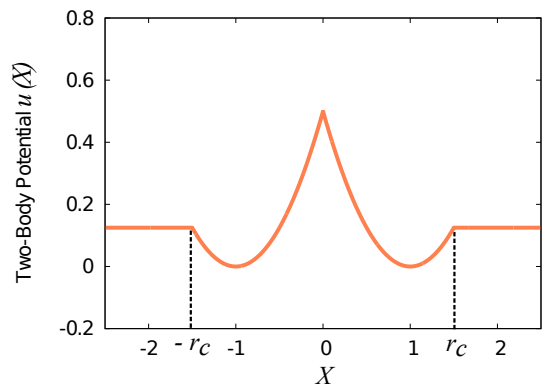}
		\caption {(Color online) Two-body potential $u (X)$ for $K = 1$, $l_0 = 1$ and $r_c = 1.5$. \label {Fig:TwoBodyPotential} }
	\end {center}
\end {figure}

We briefly review the setup to discuss the linear stability. We consider a population of proliferating cells, which are confined within the one-dimensional progenitor cell pool. Let $ \displaystyle { \{  x_k (t) \}_{k = 1}^{N (t) } } $ be the set of coordinates of the centers of them. Here $N (t)$ denotes the number of cells at time $t$. We assume that the cells obey the following overdamped Langevin equations:
\begin {equation} 
	\frac { {d} } { {d} t }  x_j (t) = - \frac {\partial}{\partial x_j} U \left ( \{  x_k (t) \} \right ) +  \xi_j (t),
	\label {SupEq:Many-bodyLangevin}
\end {equation}
for $j = 1, 2, \dots, N (t)$, where $ \xi_j (t)$'s denote independent white Gaussian noise terms satisfying $\left < \xi_j (t) \right > = 0, \left < \xi_j (t) \xi_k (s) \right > = 2 D \delta_{j, k} \delta (t - s)$ for $j, k = 1, \dots, N$ and with the noise intensity $D > 0$. $U \left ( \{ \vec x_k (t) \} \right ) $ denotes the interaction potential among neighboring cells. In the simulation described in the main text, the potential force was set to act only between nearest neighboring pairs. Here, for simplicity, we assume the following form: $  {U \left ( \{  x_k (t) \} \right ) = K \sum_{j} \sum_{k < j} u (  x_j -  x_k  ) } $, which represents stored force acting among all pairs. The two-body potential $u (X)$, which mimics the cell repulsion and adhesion forces, is assumed to include only short-range interaction with a cutoff: 
\begin {equation}
	\label {SupEq:PotentialForce}
	 - \nabla_X u (X) = - (  X  - l_0 {\rm sgn} (X)) \theta (r_c - \lvert X \rvert ) .
\end {equation}
Here, $l_0$ denotes a length scale which represents the inter-particle (cell) distance or the typical size of cells, $r_c$ denotes the cutoff length, and $\theta (x)$ is the step function. The two-body potential $u (X)$ is shown in Fig.~\ref {Fig:TwoBodyPotential}. We assume $r_c > 3 l_0 / 2$.

\subsection {Kinetics of cell fate decision}
Each proliferating cell (PC) at position $ x$ undergoes the following birth-death process 
\begin {equation}
	PC \stackrel{ }{\longrightarrow}
	\begin {cases}
		PC + PC & \hspace {0.2in} {\rm with~rate~~} \lambda w^+ ( \rho_L ( x) ) \\
		\emptyset & \hspace {0.2in}{\rm with~rate~~} \lambda w^- ( \rho_L ( x) ),
	\end {cases}
	\label {SupEq:DenDepDivision}
\end {equation}
where a new cell is created at the position $ x$ as a birth event occurs, while the cell simply annihilates as a death event occurs. The birth and death rates $w^{\pm} \left ( \rho_L ( x) \right ) $ depend on the local density of cells $\rho_L ( x; t ) $, which is defined as
\begin {equation}
	\rho_L ( x; t ) := \int d y \nu_L (  x -  y) \rho ( y),
	\label {SupEq:LocalDensity}
\end {equation}
where $\rho ( x; t) := \sum_{j = 1}^{N (t)} \delta ( x -  x_j (t))$ is the local density field and $\nu_L ( x)$ is the interaction kernel. 
%with $D_L (\vec x) = \Set {\vec y \in  \mathbb {R}^d | \lvert \vec y - \vec x \rvert < L}$. 
The parameter $L$, which we call the interaction range, expresses the length scale within which a cell can respond to change in the local density. 

\subsection {Continuum description}
We discuss the linear stability around the spatially uniform distribution of cells~\cite {Hernandez-Garcia2004, Heinsalu2010}. To this end, we consider a continuum description of the many-body Langevin equations~\eqref {SupEq:Many-bodyLangevin}. In the continuum description, the population of the cell is described by the local density field $ {\rho ( x; t) = \sum_{j = 1}^{N (t)} \delta ( x -  x_j (t)) } $. By following the argument by Dean~\cite {Dean1996} and taking into account the cell fate decision process, we obtain the dynamical equation for the local density field:

\begin {equation}
	\begin {split}
		\frac {\partial}{\partial t} & \rho ( x; t) \\
		&= D \nabla_x^2 \rho ( x; t) + K \nabla_x \left ( \rho ( x; t)  \nabla_x \Psi ( x; t) \right ) \\ & + \nabla_x \left ( \sqrt {2 D \rho ( x; t) }  \eta ( x; t) \right )\\
		& + \lambda F \left ( \rho_L ( x; t) \right ) \rho_L ( x; t) + \sqrt {\lambda \rho ( x; t)} G (\rho_L ( x; t) ) \zeta ( x; t), 
	\end {split}
	\label {SupEq:DensityLangevin}
\end {equation}
where $\nabla_x$ denotes partial differentiation with respect to $x$, and $ \eta ( x; t)$ and $\zeta (x; t)$ are white Gaussian noise fields satisfying
\begin {equation}
	\begin {split}
		\left < \eta ( x; t) \right > & = 0, \hspace {0.1in} \left < \eta ( x; t) \eta ( y; s) \right > = \delta (t - s) \delta ( x -  y) \\
		\left < \zeta ( x; t) \right > & = 0, \hspace {0.1in} \left < \zeta ( x; t) \zeta ( y; s) \right > = \delta (t - s) \delta ( x - y) 
	\end {split}
	\label {SupEq:Noise}
\end {equation}
with It\^ {o}'s forward discretization. $\Psi ( x; t)$ in the first line and $F (\rho)$ and $G (\rho)$ in the second line are given by
\begin {equation}
	\Psi ( x; t) := \int  {d} y \rho ( y; t) u ( x -  y), 
\end {equation}
\begin {equation}
	F (\rho) := \left ( w^+ (\rho ) - w^- (\rho ) \right ), 
\end {equation}
\begin {equation}
	G (\rho) := \sqrt { \left ( w^+ (\rho ) + w^- (\rho ) \right )}.
\end {equation}

\subsection {Linearaization}
We now linearize Eq.~\eqref {SupEq:DensityLangevin} with respect to $\rho ( x; t)$. Equation~\eqref {SupEq:DensityLangevin} has a nontrivial fixed point $\langle \rho ( x; t) \rangle = \rho^* > 0$. By rewriting Eq.~\eqref {SupEq:DensityLangevin} with the Fourier transform $\hat \rho (k; t)$ and expanding the density around the steady state $\hat \rho (k; t) = \rho^* \delta (k) + \delta \hat \rho (k; t)$ up to the leading order terms, 
Eq.~\eqref {SupEq:DensityLangevin} is linearized as follows:
\begin {equation}
	\label {SupEq:LinearLangevinFourier}
	\frac {\partial}{\partial t} \delta \hat \rho ( k; t) = - \Lambda (k) \delta \hat \rho ( k; t) + \hat \eta ( k; t),
\end {equation}
where the linear growth rate is defined as follows:
\begin {equation}
	\Lambda (k) := \alpha \nu_L (k) + D k^2 - K \rho^* \hat u_2 (k).
\end {equation}
Here, $\hat \nu_L (k)$ is the Fourier transform of $\nu_L (x)$, $\hat u_2 (k)$ is the Fourier transform of $u_2 (x) := \nabla^2 _x u (x)$, and $\hat \eta ( k; t)$ is defined as the Fourier transform of $\sqrt {\sigma} \zeta ( x; t) + \sqrt {\mu} \nabla_x \eta ( x; t)$, which has zero mean and the following correlation:
\begin {equation}
	\left < \eta ( k; t) \eta ( k^\prime; t^\prime) \right > = (2\pi)^2 (\sigma + \mu k^2) \delta ( k +  k^\prime) \delta (t - t^\prime).
\end {equation}
Here, the positive constants $\alpha$, $\sigma$ and $\mu$ are defined as
\begin {eqnarray} 
	\label {SupEq:Def_alpha_beta}
	\alpha & := & - \lambda F ^\prime (\rho^*) \rho^*, \nonumber \\
	\sigma & := &  \lambda G^2 (\rho^*) \rho^*, \nonumber \\
	\mu & := & 2 D \rho^*,
\end {eqnarray}
where $\alpha$ is the sensitivity of growth rate around steady state density $\rho^*$, and $\sigma$ and $\mu$ represent the amplitudes of fluctuations arising from cell birth-death kinetics and Langevin motion, respectively.

In the derivation of Eq.~\eqref {SupEq:LinearLangevinFourier}, the leading order expansions are given by 
\begin {equation}
	\begin {split}
		\lambda F (\rho_L ( x; t) ) \rho_L ( x; t) & = \lambda F^\prime (\rho^* ) \rho^* \delta \rho_L ( x; t) + O \left ( (\delta \rho)^2 \right ),
	\end {split}
\end {equation}
\begin {equation}
	\begin {split}
		\sqrt {\lambda \rho ( x; t)} G (\rho_L ( x; t) )  & = \sqrt {\lambda \rho^* } G (\rho^* ) + O \left ( \delta \rho \right ), 	\end {split}
\end {equation}
\begin {equation}
	\begin {split}
		\nabla_x \left ( \sqrt {D \rho ( x; t)}  \eta ( x; t) \right )  & = \sqrt {2 D \rho^* }  \nabla_x \eta ( x; t) + O \left ( \delta \rho \right ).
	\end {split}
\end {equation}
By defining $u_1 (x) := \nabla_x u (x)$ and $u_2 (x) := \nabla^2_x u (x)$, the two-body potential term is given by
\begin {eqnarray}
	\label {SupEq:ExpandTwoBodyPotential}
	&\nabla_x&  \left ( \rho ( x; t)  \nabla_x \Psi ( x; t) \right ) = \nabla_x \left [ \rho(x) \int dy \rho(y) u_1(x-y) \right ] \nonumber \\
	&=&  \nabla_x \rho(x) \int dy \rho(y) u_1(x-y) + \rho(x) \int dy \rho(y) u_2(x-y). \nonumber \\
	&=&  \rho^* \nabla \delta \rho(x) \int dy u_1(x-y) +  \rho^*  \int dy \delta \rho(y) u_2(x-y)\nonumber \\
 	&& + \rho^* (\rho^* + \delta \rho(x) ) \int dy  u_2(x-y) + O( (\delta \rho)^2 ), 
\end {eqnarray}
where $\rho (x;t) = \rho^* + \delta \rho (x; t)$ is used in the last equality. In the right hand side of Eq.~\eqref {SupEq:ExpandTwoBodyPotential}, the first term vanishes since $u_1 (x)$ is an asymmetric function, and the third term also vanishes since $\int d y u_2 (x - y) = [u_1 (y)]_{y = - \infty}^{+\infty} = 0$. Thus the Fourier transform of the two-body potential term is $\rho^* \hat u_2 (k) \delta \hat \rho (k; t) $ up to $O (\delta \rho)$.

Setting the noise term as zero in Eq.~\eqref {SupEq:LinearLangevinFourier}, the linear stability condition is $\Lambda (k) > 0$. For this to be satisfied at small $k$, we need $\alpha > 0$ (equivalently, $F^\prime (\rho^*) < 0$). We also assume $r_c < 3 l_0 / 2$ to guarantee that $\hat u_2 (k) < 0$.

%We have assumed that $\delta \rho (\vec y; t)$ varies slowly enough on the support of $u (\vec x - \vec y)$, so that the integral yields $ \delta \rho (\vec x; t)$ with a coefficient $\gamma = \int  {d}^d \vec y u (\vec x - \vec y)$. Therefore, the parameters in Eq.~\eqref {SupEq:DensityLinearization} are given by
%\begin {equation}
%	\begin {split}
%		\alpha = - \lambda F^\prime (\rho^*) \rho^*, \hspace {0.1in} \beta = D + \rho^* \gamma K.
%	\end {split}
%	\label {Eq:Def_alpha_beta}
%\end {equation}
%Therefore, Eq.~\eqref {SupEq:DensityLangevin_Averaged} is linearized as:
%\begin {equation}
%	\frac {\partial}{\partial t} \delta \rho (\vec x; t) = - \alpha \delta \rho_L (\vec x; t) + \beta \nabla^2 \delta \rho (\vec x; t),
%	\label {SupEq:DensityLinearization}
%\end {equation}
%where positive parameters $\alpha, \beta$ are set by Eq.~\eqref {Eq:Def_alpha_beta}.

%\subsection {Linear stability}
\subsection {Structure factor analysis}
%We are now in the position to perform the linear stability analysis. 
We employ the structure factor analysis in the linearlized Langevin system~\eqref {SupEq:LinearLangevinFourier} at steady state.
The structure factor, which quantifies the density fluctuation associated with a wavenumber $ k$, is defined as  $S ( k; t) = \left <  \lvert \delta  \hat \rho ( k; t) \rvert^2 \right >$.

\begin{figure*}[tbp]
	%\vspace {0.2in}
	\begin{center}
	\begin{minipage}{0.475\hsize}
		(a) Gaussian kernel
		\includegraphics[keepaspectratio, scale=1.0]{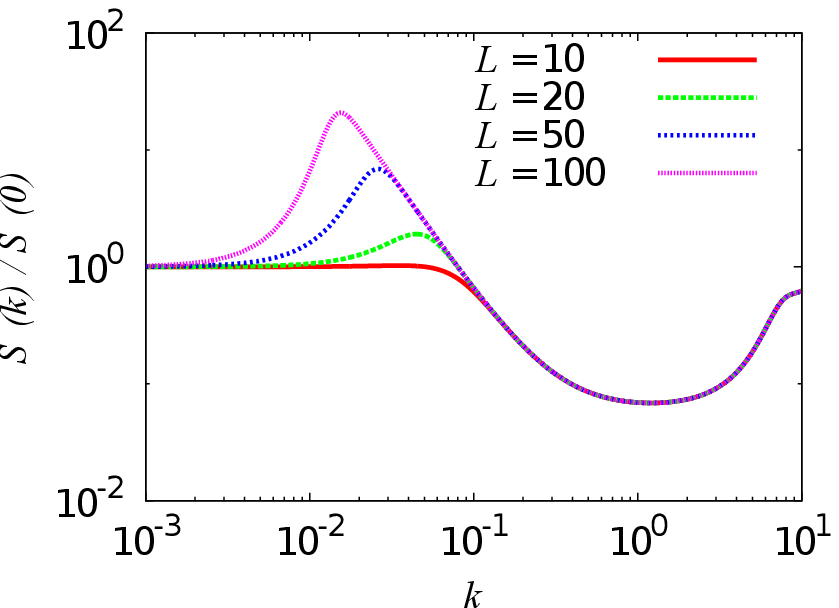}
	\end{minipage}
	\begin{minipage}{0.475\hsize}
		(b) Top-hat kernel
		\includegraphics[keepaspectratio, scale=1.0]{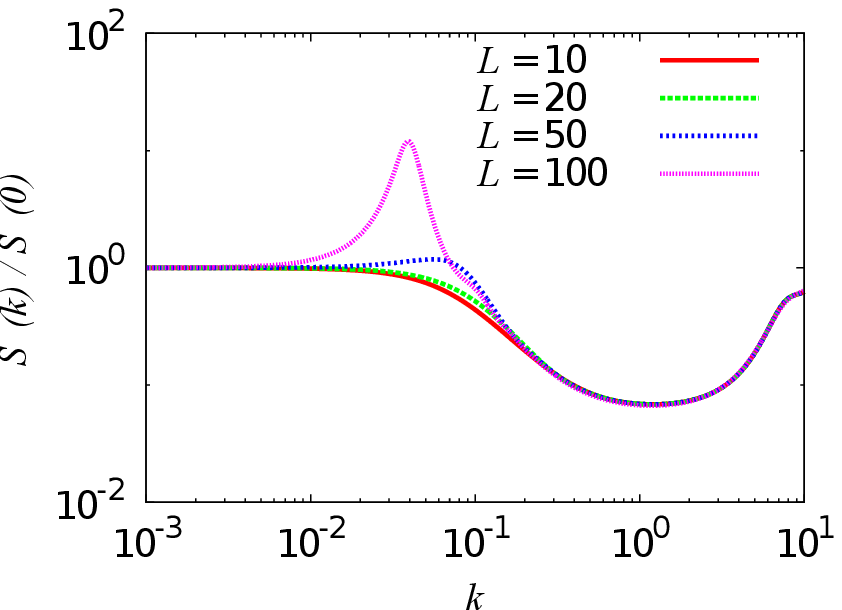}
	\end{minipage}
	\end{center}
	\caption{Structure factor $S_{\rm ss} (k)$ associated with (a) Gaussian and (b) top-hat kernel. A nontrivial peak appears after $L$ exceeds a certain threshold $L_c$. We set the parameters as $\alpha = 1$, $\sigma = 1$, $D = 5$, $\mu = 10$, $K = 1000$, $l_0 = 1$ and $r_c = 1.1$.
		\label {Fig:StrFac} }
\end{figure*}

The solution of Eq.~\eqref {SupEq:LinearLangevinFourier} with initial condition $ \delta \hat \rho ( k; t = 0) = 0$ for all $k$ is given by the following expression:
\begin {equation}
	\delta \hat \rho ( k; t) = \int_0^t d t_1 e^{ - \Lambda ( k ) ( t - t_1 )} \hat \eta (k; t_1).
\end {equation}
The structure factor is obtained as
\begin {eqnarray}
	\label {SupEq:StructureFactor}
	S ( k; t) &=& \int_0^t d t_1 \int_0^t d t_2 e^{ - \Lambda ( k) ( 2 t - t_1 - t_2)  } ( \sigma + \mu k^2 ) \delta (t_1 - t_2) \nonumber \\
	&=& \frac {\sigma + \mu k^2}{2 \Lambda ( k )}  \left ( 1 - e^{ - 2 \Lambda (k) t} \right ).
\end {eqnarray}
The structure factor $S (k; t)$ depends on the choice of interaction kernel $\nu_L (x)$. We here take Gaussian kernel $\nu_L ^{\rm G} (x) = e^{- x^2 / 8 L^2} / \sqrt { 8 \pi L^2} $ as well as top-hat kernel $\nu_L ^{\rm TH} (x) = (2L)^{-1} \theta ( L - \lvert x \rvert ) $, which is used for numerical simulations in the main text. We obtain 
\begin {equation}
	\hat \nu_L^{\rm G} (k) = e^{- 2 (kL)^2}, ~~ \hat \nu_L^{\rm TH} (k) = \frac {\sin (kL)}{kL},
\end {equation}  
respectively.
If $\Lambda ( k) > 0$ for all $ k$, the structure factor converges to the steady-state values: 
\begin {equation}
	\label {SupEq:StructureFactor_SteadyState}
	S_{\rm ss} (k) := \lim_{t \to + \infty} S ( k; t) = \frac {1}{2} \frac {\sigma + \mu k^2}{\alpha \hat \nu (kL) + D k^2 - K \rho^* \hat u_2 (k)}.
\end {equation}
If, on the other hand, $\Lambda (k)$ becomes negative for some $k$, the system is linearly unstable.

Even within the linearly stable regime, the dynamics can still result in the spatial clustering, which will show up  as a non-trivial peak in $S_{\rm ss} (k)$ (see Fig.~\ref {Fig:StrFac}). To understand the appearance of the peak, we focus on the large length scale $k \ll l_0^{-1}$. Since $\hat u_2 (k) \simeq - \gamma k^2$ for $k \ll l_0^{-1}$, where $\gamma := - \frac {d^2}{d k^2} \hat u_2 (k = 0) / 2$, the structure factor is approximated as follows:
\begin {eqnarray}
	\label {SupEq:StructureFactor_SteadyState_Aprx}
	S_{\rm ss} (k) \simeq \tilde S_{\rm ss} ( k) := \frac {1}{2} \frac {\sigma + \mu k^2}{\alpha \hat \nu (kL) + \beta k^2},
\end {eqnarray} 
where $\beta := D + K \rho^* \gamma$ is the effective diffusion constant. Note that for the potential force defined by Eq.~\eqref {SupEq:PotentialForce}, $\gamma$ is positive if $r_c < 3 l_0 / 2$, and thus $\beta > 0$.
By employing $ \tilde S^\prime (k) = 0$, we derive the condition for the existence of a peak as follows:
\begin {equation}
	\frac {2\beta}{\alpha} - \frac {2\mu}{\sigma} < \nu^{(2)} L^2,
	\label {SupEq:Condition}
\end {equation}
where $\nu^{(2)} := \int d x x^2 \nu_{L = 1} (x) > 0$. Inequality~\eqref {SupEq:Condition} naturally defines the threshold $L_c$ of interaction range, above which the spatial clustering of cells takes place:
\begin {equation}
	\label {SupEq:InteractionRangeThreshold}
	L_c = \sqrt { \frac {1}{ \nu^{(2)} } \left ( \frac { 2 \beta}{ \alpha} - \frac { 2 \mu}{\sigma} \right ) }.
\end {equation}
The phase diagram based on Eq.~\eqref {SupEq:Condition} with the top-hat kernel is shown in Fig.~\ref {Fig:Models} (c). This result implies that, even for the case of large $k$ (strong repulsion), the spatial clustering can happen if $L$ is too large. 

\subsection {Discussion}
Some studies discussed the appearance of spatial clustering by linear stability analysis~\cite {Hernandez-Garcia2004}. However, the validity of linear stability analysis depends on the choice of kernel. According to linear stability analysis, linearly unstable region appears for top-hat kernel, while it does not for Gaussian kernel. This is not a good characterization of the phenomenon. Taking into account the fluctuation of density field, and quantifying the two-point correlation of density field by structure factor, our analysis reveals that spatial clustering occurs irrespective of the choice of the interaction kernel, and specifies the parameter region where the clustering occurs. Therefore, our analysis clarifies a better understanding of the Brownian bug problem. 

\section {Numerical parameters}
We take the following functional form of $w^{\pm} (\rho)$:
\begin {equation}
	w^{\pm} (\rho) = \lambda_{\pm} + \frac {1 - e^{ \pm b (\rho - \rho_0)}}{\kappa^{-1} + \lambda_{\pm}^{-1} e^{ \pm b (\rho - \rho_0)}},
\end {equation}
with $\lambda_{\pm} = \lambda (1 \pm \Delta) / 2$, $-1 < \Delta < 1$, $b > 0$, $\kappa > \lambda$, and $\rho_0 > 0$. 
The numerical parameters are set as follows: $L_{\rm Sys} = 1000$, $K = 1$, $l_0 = 1$, $D = 0.02$, $\Delta = 0.2$, $\rho_0 = 1$, and $b = 1$. We take $\lambda = 0.025$ for Fig.~\ref {Fig:Results} (c) and (d), $\lambda = 0.005$ for Fig.~\ref {Fig:Results} (a), (b), and Fig.~\ref {Fig:CrossoverScaling}, and $\kappa / \lambda = 4$ for all cases. The discretization time step for the numerical integration of Eq.~\eqref {Eq:EOM} is set as $\Delta t = 0.01$. The clone size distribution is calculated with $10^5$ independent runs.

%%\end{abstract}
%
%% insert suggested PACS numbers in braces on next line
%%\pacs{}
%%% insert suggested keywords - APS authors don't need to do this
%%%\keywords{}
%%
%%%\maketitle must follow title, authors, abstract, \pacs, and \keywords
%%\maketitle
%
%% body of paper here - Use proper section commands
%% References should be done using the \cite, \ref, and \label commands
%%\section{}
%% Put \label in argument of \section for cross-referencing
%%\section{\label{}}
%%\subsection{}
%%\subsubsection{}
%
%
%%%

% Create the reference section using BibTeX:
%\bibliography{basename of .bib file}
%\bibliography{PRL_Ref}

\end{document}